\documentclass[usenatbib]{mnras}
\usepackage{epsfig}
\usepackage{amsmath,amssymb}
\graphicspath{{figures/}} 
\usepackage{xcolor}
\usepackage{soul}
\usepackage{aas_macros}  
\usepackage{chngcntr}
\graphicspath{{figures/}}

\newcommand{\Msolar}{\mbox{\,$\rm M_{\odot}$}}        
\newcommand{\Rsolar}{\mbox{\,$\rm R_{\odot}$}}        
\newcommand{\Lsolar}{\mbox{\,$\rm L_{\odot}$}}        

  \newcommand{\Teff}{\mbox{\,\em T$_{\rm eff}$}}         
 \newcommand{\teff}{\mbox{\,$T_{\rm eff}$}}      

  \def\simge{\mathrel{\raise1.16pt\hbox{$>$}\kern-7.0pt
    \lower3.06pt\hbox{{$\scriptstyle \sim$}}}}           
  \def\simle{\mathrel{\raise1.16pt\hbox{$<$}\kern-7.0pt
    \lower3.06pt\hbox{{$\scriptstyle \sim$}}}}           
\title[TESS photometry of extreme helium stars]{TESS photometry of extreme helium stars PV\,Tel and V821\,Cen}
\author[C. S. Jeffery, G. Barentsen, G. Handler]
       {C. Simon Jeffery$^1$\thanks{E-mail: simon.jeffery@armagh.ac.uk},
        Geert Barentsen$^2$, and Gerald Handler$^3$ \\
$^{1}$Armagh Observatory and Planetarium, College Hill, Armagh BT61 9DG, Northern Ireland\\
$^2$Bay Area Environmental Research Institute, P.O. Box 25, Moffett Field, CA 94035, USA\\
$^3$Nicolaus Copernicus Astronomical Center, Polish Academy of Sciences, ul. Bartycka 18, 00-716, Warsaw, Poland
}

\date{Accepted .....
      Received ..... ;
      in original form .....}

\pagerange{\pageref{firstpage}--\pageref{lastpage}}
\pubyear{2020}

\begin{document}

\label{firstpage}

\maketitle

\begin{abstract}
PV\,Tel variables are extreme helium (EHe) stars known to be intrinsic light and velocity variable on characteristic timescales 0.1 -- 25\,d. With two exceptions, they are best described as irregular. Light curves have invariably been obtained from single-site terrestrial observatories. We present {\it TESS} observations of two bright EHe stars, Popper's star (V821\,Cen) and Thackeray's star (PV\,Tel).  PV\,Tel is variable on timescales previously reported. V821\,Cen is proven to be variable for the first time. Neither light curve shows any evidence of underlying regularity. Implications are considered.       \end{abstract}


\section{Introduction}
Extreme helium (EHe) stars are rare early-type stars with high luminosity-to-mass ($L/M$) ratios and atmospheres virtually devoid of hydrogen \citep{jeffery11a}. Their rarity derives from being in a short-lived phase of evolution in which, probably, a carbon-oxygen and helium white dwarf merged to create a stable helium shell-burning giant which is now contracting to become, once again, a white dwarf \citep{saio02,zhang14,schwab19}. 
Their high luminosities make them unstable to strange-mode pulsations \citep{saio88b,jeffery16a}, which are of low amplitude and irregular with characteristic timescales ranging from $\approx0.5$\,d \citep[V2076\,Oph:][]{lynasgray87} to $\approx22$\,d \citep[FQ\,Aqr:][]{jeffery85a,kilkenny99c}.
Collectively, they are known as PV\,Tel variables \citep{jeffery08.ibvs}, 
although the light-curve of the class prototype remains poorly defined \citep{jones89}\footnote{Wide-area photometric surveys including superWASP and ASAS-3, as well as increasingly high-quality observations from AAVSO, have added hundreds of observations, but have not substantially improved the definition of the light curves. An analysis of these data is in preparation.}.  
An absence of evidence for variability in low $L/M$ EHe stars BD+10$^{\circ}$2179 = DQ\,Leo and HD124448 = V821\,Cen  is commensurate with the strange-mode pulsation model. 
Happenstance, both stars received variable star designations due to optimistic reports of their variability \citep{bartolini82,hill69} which were later repudiated \citep{grauer84,hill84,jeffery90}.
Variable low $L/M$ exceptions are V652\,Her and BX\,Cir which occupy a narrow range in effective temperature ($T_{\rm eff}$) over which iron-bump opacities drive regular pulsations with periods $\approx 0.1$\,d \citep{landolt75,kilkenny95,saio93}.  

Owing to the irregularity and timescales of most PV\,Tel variations, ground-based observations are beset by poor sampling in even the best cases, preventing any instructive frequency analysis or definitive characterisation of their light curves. 
No {\it bona fide} EHe star lies in a {\it Kepler} or {\it K2} field. 
Consequently, {\it TESS} has been the first instrument to obtain an uninterrupted view of any PV\,Tel variable over an interval long compared to their variations. 
This letter reports the TESS light curves of two of the best known EHe stars, HD\,124448 = V821\,Cen = Popper's star \citep{popper42} and HD\,168476 = PV\,Tel = Thackeray's star \citep{thackeray54}. 
The first is surprising, the second confirms inconclusive observations made three decades ago.

\begin{figure*}
\epsfig{file=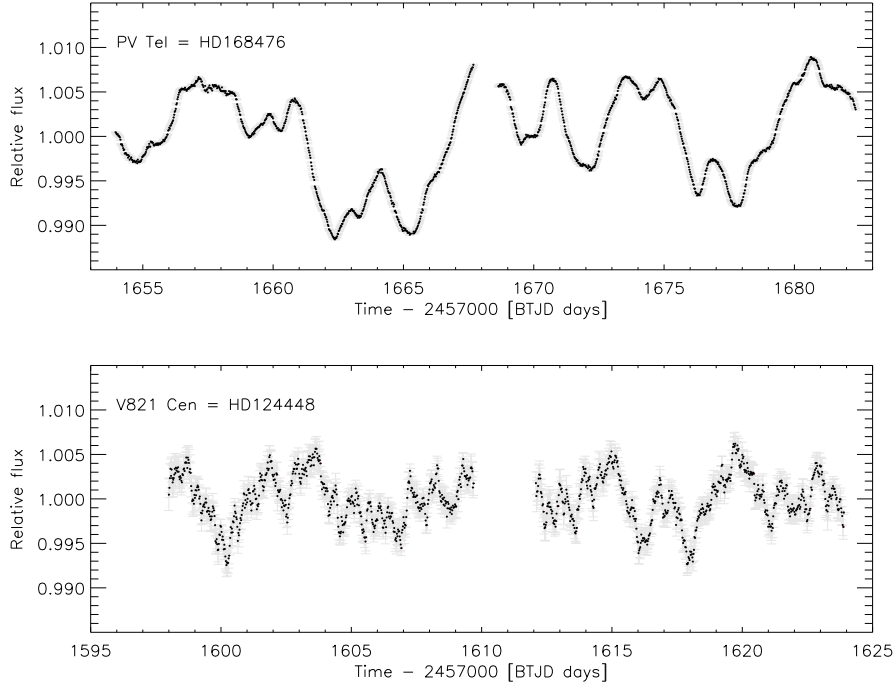,width=120mm,clip=,angle=0}
\caption{{\it TESS} light-curves for PV\,Tel and V821\,Cen. The latter has been resampled into 30\,m bins. Errors for each point are shown in light grey.} 
\label{f:lc}
\end{figure*}

\section{Observations}

The Transiting Exoplanet Survey Satellite (TESS) has a primary mission to survey the whole sky for planetary transits, staring at thousands of stars continuously for up to 40 days at a time, more if a star falls in overlapping sectors. 
This makes it ideal for discovery and follow-up photometry of all manner of variable phenomena. 

Targeted observations of V821\,Cen ($m_V = 9.98$: TIC 242416400) were made with TESS in 120\,s cadence during Sector 11, commencing 2019 April 23.
Standard data products include noise-corrected light curves which are available from the Mikulski Archive for Space Telescopes.

Observations of PV\,Tel ($m_V =  9.30$) were obtained from full-frame images obtained at the standard 30\,m cadence during Sector 13 and commencing 2019 June 19.
Instrument systematics were removed from the light curve using the {\it TessPLDCorrector} feature available in version 1.9 of the {\it Lightkurve} Python package \citep{lightkurve}.

\begin{figure*}
\epsfig{file=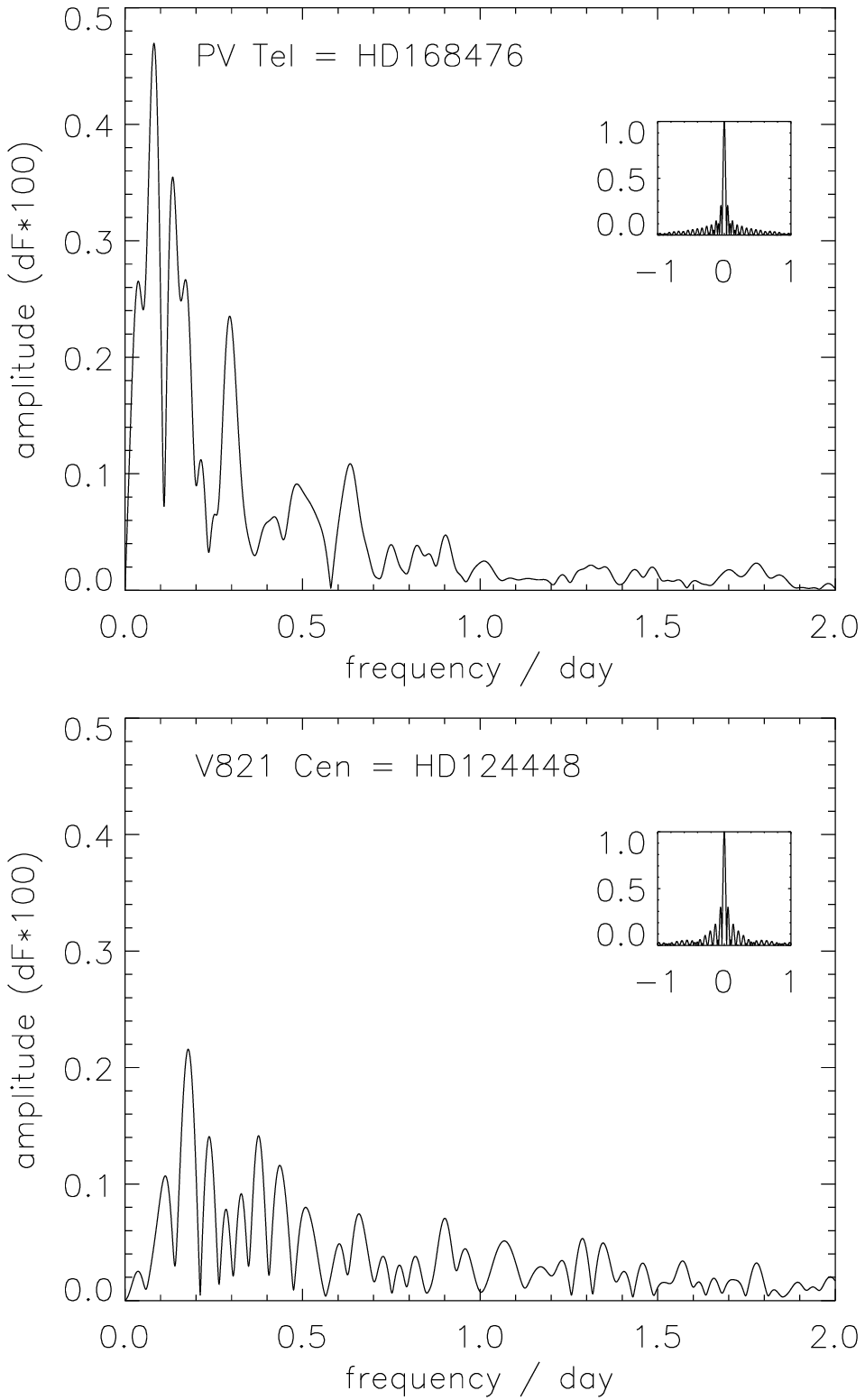,width=70mm,clip=,angle=0}
\epsfig{file=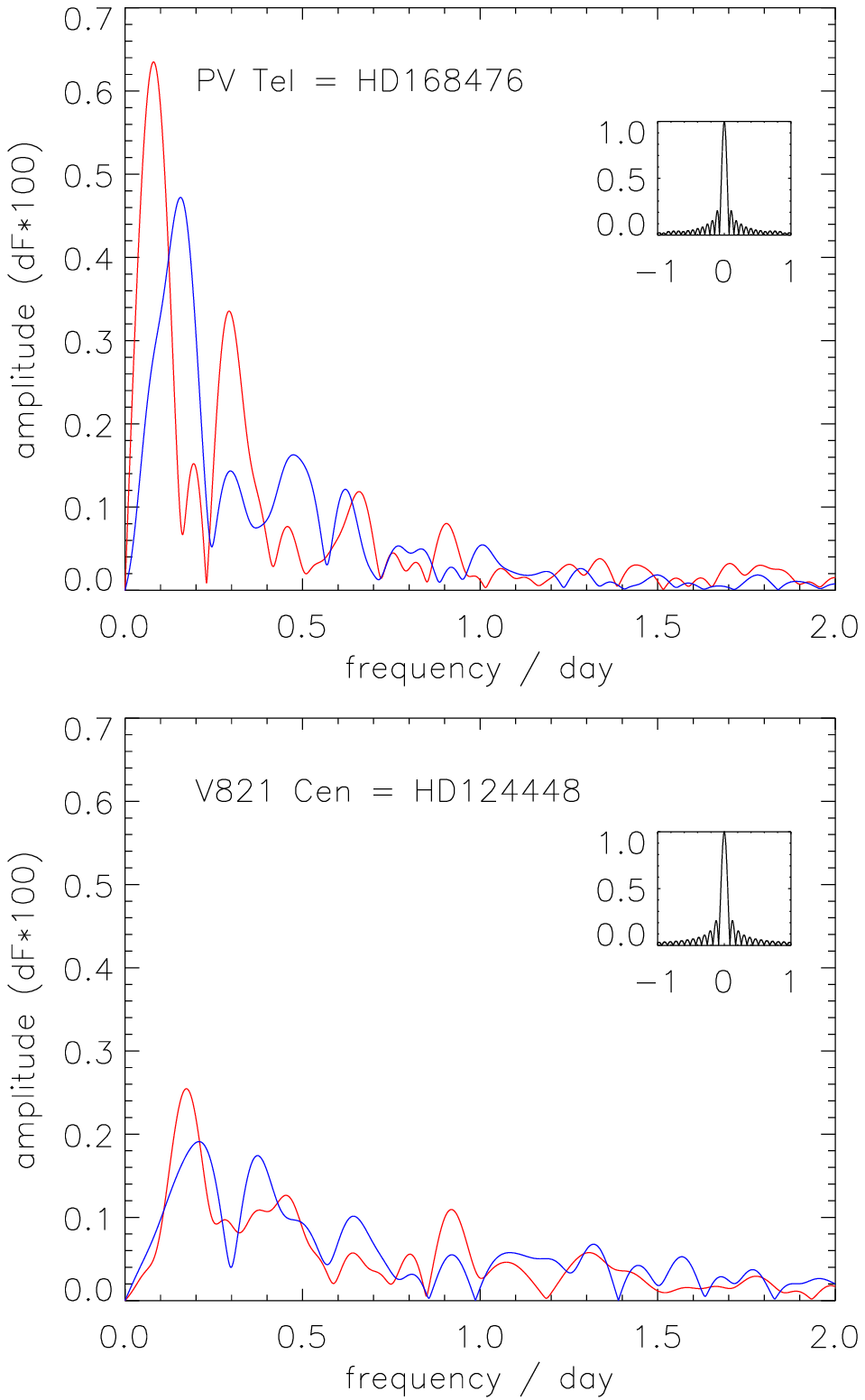,width=70mm,clip=,angle=0}
\caption{Left: part of the frequency-amplitude spectra derived from the {\it TESS} light curves for PV\,Tel and V821\,Cen. The spectral window function is inset for each case. The spectra are essentially flat for frequencies $>2$\,d$^{-1}$.  Right: As left, but for both light curves split into two equal time segments, red and blue corresponding to first half and second half segments respectively. 
}  
\label{f:ps}
\end{figure*}

\section{Results}

Figure \ref{f:lc} shows the {\it TESS} light-curves of PV\,Tel and V821\,Cen  normalized to their mean flux.  
Both are remarkable.
A part of the Fourier transforms (FT) of both light curves, in the form of amplitude spectra in the frequency range $0 - 2\,{\rm d^{-1}}$, is shown in Fig.\,\ref{f:ps}. There is no significant signal at higher frequencies. 
The width of features in the FT is inversely proportional to the lengths of the data series, being 28 and 26\,d, respectively.\\

\noindent {\bf PV\,Tel} shows variability with a full amplitude of $\approx 2.0\%$ on timescales between 1 and 5\,d.
The FT shows peaks at frequencies of 0.08, 0.13 and 0.295\,d$^{-1}$, corresponding to periods of 12.3, 7.5 and 3.4\,d. 
We identify peaks as having amplitude more than 4 times the mean power locally. 
They represent real power in the FT corresponding to a real variation in light from the star on a timescale associated with that frequency at the time the observations were made, and nothing more.    

\citet{hill69} identified microvariations in brightness and radial velocity but not in colour and found no period. 
\citet{walker85}  reported the radial velocity to be  `variable in a complex manner, possibly built up from several simple periods'.  
\citet{jones89} reported possible photometric periods in the range 8-15\,d. 
\citet{lawson93} identified a radial velocity timescale $\approx 9$\,d.  
\citet{jeffery01c} adopted a period of 6.7\,d to interpret coeval radial-velocity  and ultraviolet spectrophotometry. 
\citet{jeffery08.ibvs} adopted  8--10\,d. on the basis of the papers cited above. 

The current observations represent the first continuous clean light curve for PV\,Tel and demonstrate that the variations are unlikely to be either singly or multiply periodic. \\

\noindent {\bf V821\,cen} shows variability with a full amplitude of $\approx 1.3\%$ on timescales between 0.3 and 3\,d.
The FT shows a peak at a frequency of 0.18\,d$^{-1}$, corresponding to a period of 5.6\,d.  

\citet{landolt73} reported V821\,Cen to be variable on a timescale of days to years.
Systematic monitoring by \citet{jeffery90} placed an upper limit of 0.02 mag on any variation over timescales from hours to years. 
The current observations confirm for the first time that V821\,Cen is variable and that the amplitude is below the 0.02\,mag threshold reported previously.
As with PV\,Tel, there is little evidence for singly or multi-periodic behaviour. 

FTs were also calculated for both datasets split into two equal and non-overlapping segments.
Peaks in the FTs shown in Fig.\,\ref{f:ps} are seen in one or other but not both segments,
apart from the peak at   $\approx0.18$\,d$^{-1}$, in V821\,Cen. 
The latter has a 20\% lower amplitude in the second segment compared with the first. 
Consequently, from these data, we find little evidence for a persistent underlying period in the light curve of either star. 

\begin{table}
    \caption{Fundamental data}
    \label{t:data}
    \centering
    \begin{tabular}{ccc}
    \hline
         & PV\,Tel & V821\,Cen  \\
         \hline
    $T_{\rm eff}/{\rm K}$ & $13750\pm400$ & $16100\pm300$ \\[0.5mm]
    $\log g/ {\rm cm\,s^{-2}}$ & $1.60\pm0.25$ & $2.30\pm0.25$ \\[0.5mm]
     & \multicolumn{2}{c}{\citet{pandey06}} \\[2mm]
    $\log L/M / (\Lsolar/\Msolar)$ & $4.34\pm0.25$ & $3.92\pm0.25$ \\[0.5mm]
    $\log \bar{\rho}/\rho_{\odot}\,^{\star}$  & $-4.16\pm0.38$ & $-3.11\pm0.38$ \\[0.5mm]
    $P_{\rm F}/{\rm d}\,^{\star,\ast}$ & $4.8\pm2.1$ & $1.4\pm0.6$ \\[0.5mm]
     & \multicolumn{2}{c}{by identity} \\[2mm]
    $m_G$ & 9.23 & 9.94 \\[0.5mm]
    $\pi /{\rm mas}$ & $0.12\pm0.04$ & $0.57\pm0.13$ \\[0.5mm]
         & \multicolumn{2}{c}{\citet{gaia18.dr2}} \\[2mm]
    $d /{\rm kpc}$ & $7.1^{+1.4}_{-2.0}$ & $2.04^{+0.43}_{-0.71}$ \\[0.5mm]
    $ \log L/\Lsolar $ & $4.38^{+0.13}_{-0.23}$ & $3.10^{+0.13}_{-0.31}$ \\[0.5mm]
    $ R/\Rsolar$ & $27.20^{+4.09}_{-7.26}$ & $4.57^{+0.70}_{-1.63}$ \\[0.5mm]
     & \multicolumn{2}{c}{Martin \& Jeffery (in prep.)} \\
     \hline
     \multicolumn{3}{l}{${\star}$: assuming mass $M=0.8\Msolar$, $\ast$: assuming $P_{\rm F} \sqrt{\frac{\bar{\rho}}{\rho_{\odot}}} \approx 0.04\,{\rm d}$ }  \\
    \end{tabular}

\end{table}

\section{Discussion}

Since discovery of a 21\,d sinusoidal variation in FQ\,Aqr \citep{jeffery85a}, the question has been how to characterise the PV\,Tel variables. 
Accruing data has so far failed to find a regular period in stars other than V652\,Her and BX\,Cir. 
They must therefore be described as irregular, but whether random or chaotic is unresolved.
The question is moot since the latter can be described by non-linear dynamics and are therefore (at least partially) deterministic. 
Phase delay diagrams fail to demonstrate order for either light curve over any delay time, so we are forced to conclude, at least from these data, that PV\,Tel variations are  random. 

Table\,\ref{t:data} shows fundamental and associated data for both programme stars, including effective temperature \teff, surface gravity $g$, $L/M$ ratio, mean density $\bar\rho$ for an assumed mass, pulsation period $P_{\rm F}$ assuming the classical period-mean density relation \citep[][Eq. 26]{eddington18},  Gaia magnitude $m_{\rm G}$, parallax $\pi$, distance $d$, luminosity $L$ and radius $R$. 
Derived masses are rendered meaningless by the combination of errors in distance and gravity and are not reported.
For any reasonable assumption about their masses, V821\,Cen has a significantly higher mean density and hence a theoretical pulsation timescale shorter than PV\,Tel. 
Whilst a qualitatively shorter characteristic timescale for V821\,Cen might be inferred from Fig.\,\ref{f:lc}, a quantitative demonstration remains elusive.


Two related hydrogen-deficient stars were observed with {\it K2}.
One was  the relatively hydrogen-rich and metal-poor HD\,144941 \citep{harrison97,jeffery97}, which shows a periodic light curve best interpreted by a rotational modulation  due to starspots \citep{jeffery18}.
The second was the hot R\,CrB star V348\,Sgr which shows highly irregular variations. 
Although a cluster of frequencies  0.2 and 0.4\,d$^{-1}$ and  isolated peaks at 1.6, 2.4, and 3.2\,d$^{-1}$ are evident in the FT,  these do not persist throughout the observing cycle \citep{jeffery19a}.   
However, there are similarities between the {\it K2} light curve of V348\,Sgr and the {\it TESS} light curve of V821\,Cen, with variability evident on multiple timescales. 
V348\,Sgr has $\Teff = 22\,000\pm2\,000\,{\rm K}$ and $\log L/M / (\Lsolar/\Msolar) < 4.06$ \citep{jeffery95a}. 
Being significantly hotter than V821\,Cen, but with a similar $L/M$ ratio, the mean density $\log \bar{\rho}/\rho_{\odot}>-2.5\pm0.4$ is substantially higher. 
Variability related to pulsation should therefore occur on shorter timescales as observed. 

EHe's have been associated with the cooler R\,CrB variables.
Both classes are hydrogen-deficient and carbon-rich \citep{asplund00,jeffery08c}. 
It has been proposed that the characteristic deep minima seen in R\,CrB stars are due to carbon-rich material ejected from the surface in the line of sight \citep{loreta35,okeefe39,clayton96} and that the ejection mechanism is associated with pulsation \citep{feast86,woitke96}.
Efforts to detect pulsation have, so far, only uncovered small-amplitude light variations similar in character to those reported previously in EHe stars
\citep{alexander72,jones89}.

\citet{jeffery08.ibvs} subdivided the PV\,Tel variables into three types. 
Type III includes the regular pulsators V652\,Her and BX\,Cir. 
Type II includes hot variables which vary on timescales too long to be associated with the radial fundamental mode. 
PV\,Tel itself and V821\,Cen belong to Type I, which includes variables at the cool end of the EHe temperature sequence, 
and which show characteristic timescales consistent with fundamental radial mode pulsation  \citep{saio88b}. 
In view of the very high $L/M$ ratios of these stars, the envelopes are very extended and include a density inversion. 
Pulsations are extremely non-adiabatic and are likely to be non-linear and possibly  not spherically symmetric. 
Hydrodynamic models which fully reproduce the observed behaviour have never been completed, although partial successes have been achieved \citep{willingale90.thesis,montanes02.thesis}.
Analogies with the structure and behaviour of other luminous and irregular pulsating stars such as the W\,Vir and RV\,Tau type variables will be instructive.  
Such work may help solve problems relevant to a range of astrophysical phenomena, including the origin of R\,CrB activity.  

\section{Conclusion}

We have presented {\it TESS} light curves for two extreme helium (EHe) stars. 
One was a known PV\,Tel variable with a poorly defined light curve.
The second was previously considered non-variable. 
These are emphatically the best light curves ever obtained for any Type I PV\,Tel variable. 
Both are clearly variable on characteristic timescales of a few days. 
The variations are irregular but more likely to be random than chaotic, which presents difficulties for establishing a model. 
The timescales are consistent with pulsation in the fundamental radial mode as obtained from linear non-adiabatic analyses; however such analyses cannot address the lack of regularity. 
Future work should aim to characterise the variability such that each star's properties can be compared systematically. For example: are 
characteristic timescales associated with radius? Are amplitudes associated with $L/M$ ratio? Can these variations explain the R\,CrB phenomenon?

\section*{Acknowledgments}
This paper includes data collected by the TESS mission. Funding for the TESS mission is provided by the NASA Explorer Program.

\bibliographystyle{mnras}
\bibliography{ehe}
\label{lastpage}
\end{document}